# Interfacial anisotropic exciton-polariton manifolds in ReS$_2$

Devarshi Chakrabarty[†1], Avijit Dhara[†1], Kritika Ghosh[2], Aswini K Pattanayak[1], Shreyashi Mukherjee[1], Ayan Roy Chaudhuri[2], Sajal Dhara*[1]

[1]Department of Physics, Indian Institute of Technology Kharagpur, Kharagpur, 721302, India

[2]Materials Science Centre, Indian Institute of Technology Kharagpur, Kharagpur, 721302, India

*Corresponding Author: sajaldhara@phy.iitkgp.ac.in

[†]These authors have contributed equally

**Abstract**

Light-matter coupling in van der Waal's materials holds significant promise in realizing Bosonic condensation and superfluidity. The underlying semiconductor's crystal asymmetry, if any, can be utilized to form anisotropic half-light half-matter quasiparticles. We demonstrate generation of such highly anisotropic exciton-polaritons at the interface of a biaxial layered semiconductor, stacked on top of a distributed Bragg reflector. The spatially confined photonic mode in this geometry couples with polarized excitons and their Rydberg states, creating a system of highly anisotropic polariton manifolds, displaying Rabi splitting of up to 68 meV. Rotation of the incident beam polarization is used to tune coupling strength and smoothly switch regimes from weak to strong coupling, while also enabling transition from one three-body coupled oscillator system to another. Light-matter coupling is further tunable by varying the number of weakly coupled optically active layers. Our work provides a versatile method of engineering devices for applications in polarization-controlled polaritonics and optoelectronics.

**Introduction**

Exciton-polaritons generated by light-matter coupling in solid state systems have been a subject of intense research, revealing new pathways to many-body phenomena such as Bose-Einstein condensation, superfluidity and polariton lasing, while also holding significant promise in myriad optoelectronic applications [1–6]. These half-light, half-matter particles formed by the hybridization of excitonic and the photonic states have been investigated earlier in semiconductor quantum well systems [7–10], and more recently in two-dimensional monolayer transition metal dichalcogenides (TMDs) [11–16]. This family of materials has provided a promising platform for exploring novel phenomena in two dimensions such as valley polarized polaritons [17–21], anisotropic polaritons [22–25], and more recently,

Rydberg state exciton-polaritons (REPs) with enhanced nonlinear optical properties [26–28]. Group VII TMDs like ReS$_2$ support excitons with binding energies of ~100 meV even in their bulk form [29]; these excitons show unique anisotropic optical and electronic properties due to the reduced crystal symmetry of the material [30–35]. Two additional excitons X$_3$ and X$_4$ were reported earlier along with the two highly polarized X$_1$, X$_2$, and their Rydberg states in ReS$_2$ [36]. More recently, splitting of X$_1$ and X$_2$ exciton peaks has been reported in ReS$_2$ placed on a gold mirror [24] which were attributed to the self-hybridized polariton modes by the authors. However, it is unclear if these modes reported are polariton modes or the previously reported X$_3$ and X$_4$ peaks which were observed even for 11 nm thin ReS$_2$ layer where there is no self-hybridized mode [36] and also on any substrate even without a bottom gold mirror.

In this work, we demonstrate for the first-time strong light-matter interaction with clear anti-crossing and Rabi splitting of up to ~68 meV in an exciton manifold which is selectively tunable using the polarization of incident excitation, by utilizing the interaction of excitons and the supported photonic mode in the interface of ReS$_2$ placed on a DBR structure. Our experiment and simulation results show that the DBR underneath the optically active layer plays an important role in enhancing the photon lifetime inside the material resulting in the formation of interfacial exciton-polaritons, which inherit polarization dependent properties from the excitons in ReS$_2$. Polaritons formed in these materials are likely to show unique, polarization-dependent properties, opening up possibilities for novel applications in polarization-sensitive switches, polarization tunable polariton lasers [37,38] and engineering topologically nontrivial polariton dispersion [39]. Strong-light matter coupling has also been demonstrated as surface Bloch wave polaritonic states at the interface of a monolayer and a DBR [40]. Such geometries offer easy access to interface the optically active material for optoelectronic applications and probing quantum many-body physics. In addition, theoretical studies have predicted that low symmetry and strong anisotropic materials are most suitable for exhibiting superfluidity [41].

The advantage of ReS$_2$ is that the material's optical response remains unaffected with the layer thickness due to its weak interlayer coupling [42], which provides ample opportunity to engineer strong-light matter coupling by altering the thickness of the top ReS$_2$ layer. By varying the in-plane polarization of incident light, it is possible to switch the system between two independent sets of three-body coupled oscillators consisting of two different exciton-polariton manifolds. Such tunability allows the coupling strength *g* to change continuously to switch regimes from weak to strong coupling, and vice versa. We find two independent coupled oscillator systems with different in-plane polarizations existing on the same platform, each linked to the two prominent exciton-polariton species along with their Rydberg states. Furthermore, we observe the formation of Rydberg state exciton-polaritons (REPs), which have recently drawn attention due to their enhanced exciton-exciton interaction leading to polariton blockades [27]. Finally, we explore the possibility of temperature tuning as an additional knob to control the coupling strength, by observing the temperature evolution of the polariton dispersion of this intriguing exciton-polariton manifold.

**Distributed Bragg Reflector supported anisotropic polaritons**

As shown in Figure 1a, alternating layers of $SiO_2$ and $Ta_2O_5$ on a Si substrate are deposited using sputtering technique to form a 10-pair (20 layers) DBR with stopband centered around wavelength $\lambda_c$ (775 nm). This ensures all excitonic resonances present in $ReS_2$, as well as their Rydberg excitations, would lie inside the high-reflectance stopband. $ReS_2$ flakes are mechanically exfoliated from commercially available bulk $ReS_2$ crystal and dry-transferred on the predefined position on the DBR. The inset shows the optical microscope image of one such 120 nm thick $ReS_2$ flake (see Figure S1 in Supplemental Material for AFM measurement). The transverse electric (TE) mode electric field profile inside the structure for $\lambda$=760 nm is overlaid on the schematic, demonstrating how interference in the $ReS_2$/DBR structure increases the electric field intensity in the optically active top layer of $ReS_2$. For the range of angles probed by the 0.7 NA objective lens used in our measurement setup, the transverse magnetic (TM) mode electric field has no significant difference and practically overlaps with the TE mode, and is therefore not separately considered. The photonic mode present inside the energy range of the DBR stopband is reminiscent of optical Tamm states [43–45], but we make a distinction here since the top $ReS_2$ layer is a semiconductor with exceptionally high refractive index [46] instead of a metallic layer. The $ReS_2$ layer on top of a DBR thus creates a highly dispersive photonic mode with narrower linewidth than Fabry–Pérot modes in free standing $ReS_2$, which is capable of inducing strong light-matter coupling and thus forming exotic exciton-polaritons which inherit the unidirectional anisotropic properties of the parent excitons.

The signature of highly anisotropic exciton-polaritons was at first probed via polarization-resolved differential reflectance and photoluminescence (PL) at 3.2K, the results of which are shown in Figure 1b and 1c respectively, where the polarization angle $\theta_P$ of linearly polarized incident light is w.r.t the b-axis of $ReS_2$. Incident light is focused within a sub-micrometer spot on the sample plane using a 0.70 NA objective. The input port contains a linear polarizer followed by a half wave plate mounted on a motorized rotational stage in order to control the polarization of incident light. For polarization-resolved PL measurements, a 660 nm CW laser is used, with an analyzer kept in the output path.

Interestingly, we observe that absorption and PL at otherwise weak exciton resonances ($X_3$ and $X_4$), which were reported in our earlier work in several-layer $ReS_2$ exfoliated on $SiO_2$/Si substrate [36], are now strongly enhanced due to light matter coupling. Absorption maximum for ($X_1$, $X_3$) and ($X_2$, $X_4$) appears at $\theta_P = 3°$ and 78° respectively, agreeing with previously reported values of their transition dipole moment orientation [30,47]. However, the shape of the absorption dip of excitons $X_1$ and $X_2$ in the color plot show a significant departure from that of $ReS_2$ on $SiO_2$ substrate (a comparison plot is included as Figure S2 in the Supplemental Material). This can be understood subsequently from the angle resolved polariton dispersion, which shows an incident polarization dependence. Strong light-matter coupling around $\theta_P = 3°$ (78°), for which the exciton oscillator strength of $X_1$ ($X_2$) is maximized, leads to the lower polariton branch (LPB) acquiring a large dispersion. This manifests as increased linewidth and a red-shift in integrated reflectance. The absorption dips from all four excitons, as well as the LPBs can be seen in the bottom panel of Figure 1b, which corresponds to differential reflectivity for $\theta_P = 130°$. We observe the absorption from the first and second Rydberg excitations (denoted subsequently as $X_i^{(n=2)}$ and $X_i^{(n=3)}$, $i=1,2$) of $X_1$ and $X_2$ is greatly enhanced due to coupling with the overlapping photonic mode and formation of REPs. The PL as seen in Figure 1c is similar to the reflectivity data with all the excitonic features including the Rydberg states (also see polar plot Figure S3 in the Supplemental Material). The PL intensity is dominated by the LPB corresponding to each particular analyzer angle $\theta_A$. The long tail in the reflectivity and PL data below 1.64 eV is attributed to the highly dispersed middle polariton branch (MPB).

Interestingly, faint absorption dips as well as PL are observed at the unperturbed energies of excitons $X_1$ and $X_2$ above the lower polariton branch, indicating a fraction of excitons are weakly coupling with light. This is because confined photon modes are in the lower-half plane (LHP), defined as the dielectric stack below the air-$ReS_2$ interface, and are strongly coupled to the material, while there also exist unconfined continuum photon modes in the upper-half plane (UHP) above $ReS_2$. The LHP modes mix with a fraction of excitons in $ReS_2$, primarily in the layers which lie in the antinodes of the electric field. The rest of the excitons are still weakly coupled to the UHP modes, so they appear in the UHP reflection and PL.

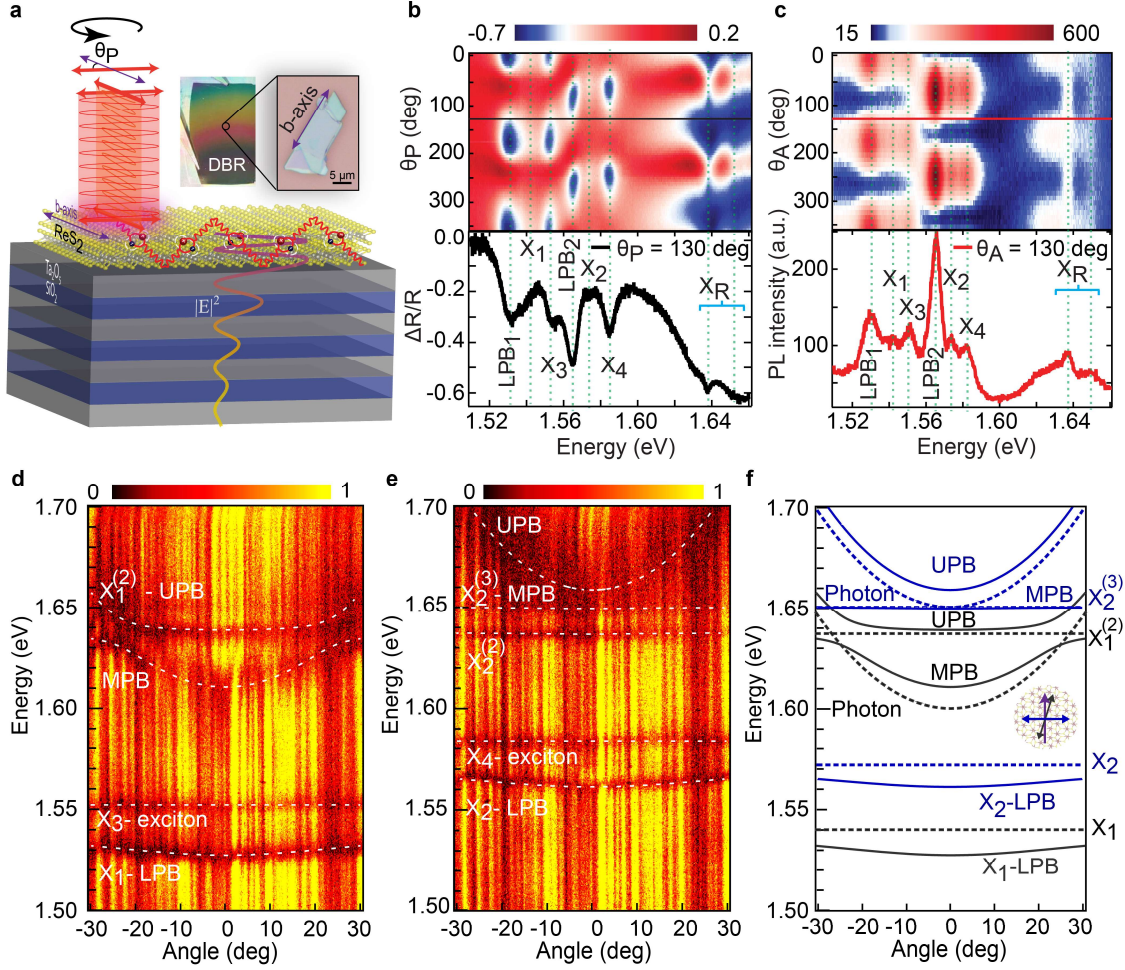

**Fig. 1. Polarization-resolved optical measurements revealing exciton-polaritons in ReS$_2$.** (**a**) Schematic of the ReS$_2$/distributed Bragg reflector (DBR) structure. (Top right) Optical microscope image of a 120 nm thick ReS$_2$ flake transferred on the DBR, showing preferential cleavage along the crystal's b-axis. Linearly polarized white light is used for reflectivity measurements, with polarization angle $\theta_P$ with the b-axis. The transverse electric field profile for TE mode is overlaid on the structure, representing a photonic mode within the DBR stopband which can couple with excitons. (**b**) Color plot representing the differential reflectance spectrum as a function of $\theta_P$ for the 120 nm ReS$_2$/DBR device. Line profile in the lower panel shows reflectance at $\theta_P = 130°$, where absorption dips from all four excitons species and the two prominent lower exciton-polariton branches can be observed. X$_R$ denotes the Rydberg excitations. (**c**) Color plot showing the PL spectrum as a function of angle of the analyzer $\theta_A$ with respect to b axis. PL spectrum at $\theta_A = 130°$ is shown in the lower panel. The subscript 1 or 2 added to the polariton branch labels indicate association with X$_1$ or X$_2$ excitons. (**d, e**) Experimental angle resolved reflectance for two different incident polarization angles ($\theta_P = 170°$ and $90°$), corresponding to two different polariton manifolds associated with X$_1$ and X$_2$ respectively. Dashed white line used as a guide to the eye to show the polaritons branches. $X_1^{(2)}$ and $X_2^{(3)}$ denotes the first and second excited Rydberg states of X$_1$ and X$_2$ respectively. (**f**) Dispersion relations of the uncoupled excitons and photonic mode (dashed lines), and coupled polariton modes (solid lines). Black and blue color is used to identify two different three-body coupled polariton manifolds corresponding to angle $\theta_P = 170°$ and $90°$ respectively.

The intriguing energy dispersion of the exciton-polariton manifolds was probed via single-shot imaging of the Fourier plane of the objective lens [15]. Figures 1d and 1e show angle-resolved reflectivity spectra at 3.2 K for the two cases, when incident light polarization is aligned to excite only X$_1$ and X$_2$ respectively ($\theta_P = 170°$ and $90°$), thus creating two different exciton-polariton manifolds. The lowermost absorption dips in both cases show a pronounced parabolic dispersion, which are identified as $X_1$-LPB and $X_2$-LPB. The upper polariton branch (UPB) in Figure 1d and MPB in Figure 1e are denoted as $X_1^{(2)}$- UPB and $X_2^{(3)}$- MPB respectively. Figure 1f shows the polariton branches (solid black for $\theta_P = 170°$ and solid blue lines for $\theta_P = 90°$) obtained from the two independent three-body coupled oscillator models fitted with experimental data, which is discussed subsequently in detail.

The two associated shoulder peaks $X_3$ and $X_4$ show negligible dispersion in this system despite having significant oscillator strength, indicating an inherently lesser propensity to couple with the in-plane electric field. This could mean their dipole moments are partially oriented out-of-plane. The origin of $X_3$ and $X_4$ as also reported in our earlier work on 11 nm $ReS_2$ [36] still requires further theoretical investigation. We note that it is not possible to explain these additional peaks as self-hybridized polariton modes as reported in a related work [24], since simulating their response required adding separate Lorentz oscillators to the permittivity model.

**Polarization tunable dispersions in exciton-polariton manifolds**

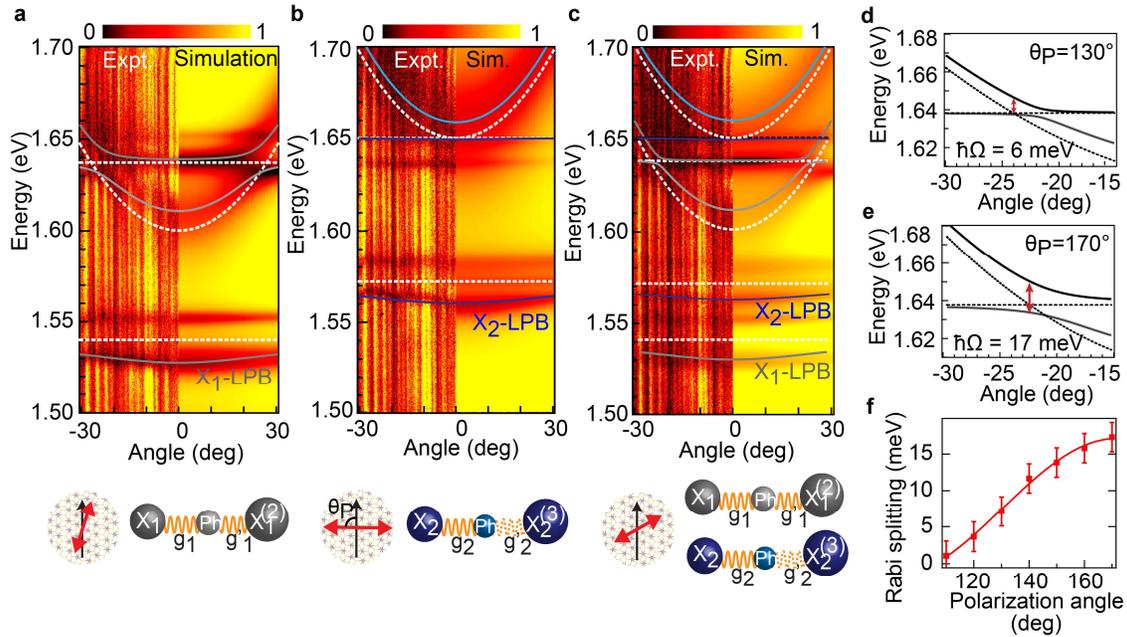

Fig. 2. Polarization-controlled light-matter coupling. (a) Angle-resolved reflectance of 120 nm $ReS_2$ on DBR for 3.2K (left) compared with transfer matrix simulation (right) when the incident linear polarized light is aligned at $\theta_P = 170°$, so that only the $X_1$ exciton-polariton manifold is excited. Dispersions of the uncoupled photon mode and excitons (white dashed lines) and the exciton-polariton modes (grey solid lines) are indicated. The spring and mass diagram below represents the corresponding three-body coupled oscillator system involving the photon mode (Ph), $X_1$ and the first Rydberg excitation $X_1^{(2)}$. $g_1$ and $g_1'$ are the coupling constants. (b) Angle-resolved reflectance of the system when $\theta_P = 90°$, such that incident polarized light excites only $X_2$. The coupled oscillator system here consists of the photon mode for this polarization (Ph), $X_2$ and the second Rydberg excitation $X_2^{(3)}$. $g_2$ and $g_2'$ are the corresponding coupling constants. Springs drawn using dashed lines

indicate weak coupling. Exciton-polariton modes for this polarization are indicated with blue solid lines. (**c**) Angle-resolved reflectance for $\theta_P$ = 130°, where both the coupled oscillator systems shown in (a) and (b) are present, but with changed coupling strength. (**d, e**) shows the Rabi splitting between $X_1^{(2)}$- UPB and MPB ranging from 6 meV to 17 meV as the polarization angle is changed from 130° to 170°. (**f**) Rabi splitting from (d, e) as a function of polarization angle $\theta_P$. Solid red line indicates a sinusoidal dependence of coupling with $\theta_P$.

Simulation obtained from the transfer matrix method is plotted alongside the experimental angle-resolved data for two different polarization angles ($\theta_P$ = 170° and 90°) in Figure 2a and 2b (see Notes 1 and 2 in the Supplemental Material for details on polariton dispersion fitting, and Figures S4-6 for the reflectivity cross-sections used to extract polariton dispersions). Fitting this with the experimental data, we obtain the background permittivity $\epsilon_b$ to be 23 and 18.5 along the $X_1$ and $X_2$ orientations respectively, which is in good agreement with theoretically predicted values for the bulk crystal [46].

At each polarization, the photonic mode supported in the semiconductor layer (dashed white line with parabolic dispersion) is modified due to the anisotropic background permittivity. We observe three prominent polariton modes in both cases, resulting from coupling between the photonic mode ($E_{Ph_i}$), the strongest exciton species ($X_i$), and one of its Rydberg excitations ($X_i^{(n)}$). Thus, for these two polarizations, we model the system using the three-body coupled oscillator Hamiltonian [8]:

$$\begin{pmatrix} E_{Ph_i}(\hbar\omega) + i\hbar\Gamma_{Ph_i} & g_i & g_i' \\ g_i & E_{Xi} + i\hbar\Gamma_{Xi} & 0 \\ g_i' & 0 & E_{X_i^{(n)}} + i\hbar\Gamma_{X_i^{(n)}} \end{pmatrix}$$

Here, the subscript $i$ =1, 2 denotes $X_1$ and $X_2$ orientations respectively. $E_{Ph_i}(\hbar\omega)$ is the dispersion of the corresponding bare photonic mode for each orientation, which we have obtained experimentally using data taken at room temperature (see Figure S7 and Note 3 in the Supplemental Material). $\hbar\Gamma_{Ph_i}$ and $\hbar\Gamma_{Xi}$ are the HWHM of the photonic mode and excitons respectively. The exciton halfwidths were typically 2 meV at 3.2K for this system. $\hbar\Gamma_{Ph_i}$ was found to be 6 meV. Note that we neglect exciton-exciton coupling here. We also neglect coupling of the $X_3$ and $X_4$ excitons with the photonic mode, since they show negligible dispersion experimentally. Diagonalizing this matrix yields the polariton mode dispersions, which can be fitted with experimentally observed dispersions to obtain the coupling strengths. For $X_1$ polarization, we obtain coupling strengths $g_1$ = 30 meV and $g_1'$ (coupling with the first Rydberg excitation $X_1^{(n=2)}$) = 7.5 meV, satisfying the condition for strong light-matter coupling [13] ($g > \hbar\sqrt{\Gamma_{Ph_i}^2 + \Gamma_{Xi}^2}$ = 6.3 meV) for both interactions. Rabi splitting of 17 meV is observed between the upper and middle polariton branches. For $X_2$ polarization, we obtain coupling strengths $g_2$ = 32 meV, and $g'_2$ (coupling with the second Rydberg excitation $X_2^{(n=3)}$) = 3 meV. The former falls in the strong coupling regime whereas the coupling with $X_2^{(n=3)}$ is weak, likely due to the low oscillator density of the higher excited (n=3) exciton state.

Figure 2c shows reflectivity when the incident polarization angle is between $X_1$ and $X_2$'s orientations, specifically for $\theta_P$ = 130°. Since the triclinic crystal structure of $ReS_2$ renders it an optically anisotropic biaxial medium, the incident electric field propagating in the z-direction (into the sample) gets decomposed into two components along the two optical axes, each of which has its own dielectric permittivity [48], the values of which we have obtained earlier. Thus, at this polarization angle, both the coupled oscillator systems in Figure 2a and 2b are excited, independently of each other. As evidence of this we see two different coupled photonic modes associated with both polarization orientations, along with $X_1$-LPB and $X_2$-LPB, appear in the reflectivity spectra. The coupling strength has the following dependence:

$$g \propto \sqrt{n * f / V_m} \quad (1),$$

where $n$, $f$, and $V_m$ are number of excitons, oscillator strength per exciton, and mode volume respectively [10]. $n$ can be increased by increasing the number of decoupled layers of $ReS_2$. Since $f$ is proportional to the squared modulus of the transition dipole moment, for linearly polarized excitons we get $f \propto cos^2(\theta_x - \theta_P)$, where $\theta_x$ is the angle along which the excitons are polarized. From (1), we find the modulation of $g$ is therefore a consequence of the change in incident polarization $\theta_P$, which becomes a convenient knob to tune the coupling strength. By fitting the experimental data with two independent three-body coupled oscillator systems, we find the coupling strengths $g_1$ and $g_2$ are both reduced to 22 meV. This agrees with the expected values $g_1^{X1,X2} = g_1^{X1} cos(170° - 130°) = 23.0$ meV and $g_2^{X1,X2} = g_2^{X2} cos(90° - 130°) = 24.5$ meV, where the superscripts are used to denote the corresponding orientation of incident polarization.

As a result of reduced coupling, the dispersion of the LPBs is also visibly reduced, which manifested as a modulation of the absorption dip linewidth in the integrated reflectance shown in Figure 1b. $g'_1$ falls to 2.5 meV, entering the weak coupling regime, while $g'_2$ becomes 2 meV. The transition from weak to strong coupling is evident in the appearance of Rabi splitting for the REP, which is difficult to resolve as we turn the incident linearly polarized light away from the dipole moment orientations of $X_1$ or $X_2$ (Figure 2d), but becomes prominent in the strong coupling regime (Figure 2e). Figure 2f shows how the experimentally obtained Rabi splitting increases in a sinusoidal fashion as the incident polarization angle is tuned toward the dipole orientation of $X_1$, consistent with the fact that the projection of the electric field on the dipole moment is being tuned. Thus, the coupling strength for both species of excitons can be smoothly tuned from their minimum to maximum values by changing the incident polarization.

**Thickness-controlled detuning and Rabi splitting**

Next, we demonstrate the effect of detuning in the formation of exciton-polaritons in this system, by varying the thickness of the $ReS_2$ layer and the central-wavelength at various position on the DBR structure with thickness gradient of the dielectric layers (as shown in Figure 1a). Since the energy range of the photonic mode is determined by the thickness of the top $ReS_2$ layer, the $ReS_2$/DBR system can be easily utilized to achieve zero or negative detuning between the exciton resonance and photonic mode. The ability of thickness tuning to enhance absorption and light-matter coupling is captured clearly in the transfer-matrix simulation, shown as Figure S8 in the Supplemental Material. Angle-resolved reflectance measurements were also performed on $ReS_2$ flakes of different thickness deposited on the same DBR. Figure 3a shows the angle-resolved reflectance spectrum for a 145 nm $ReS_2$ flake on DBR, with incident light polarization aligned along $X_1$.

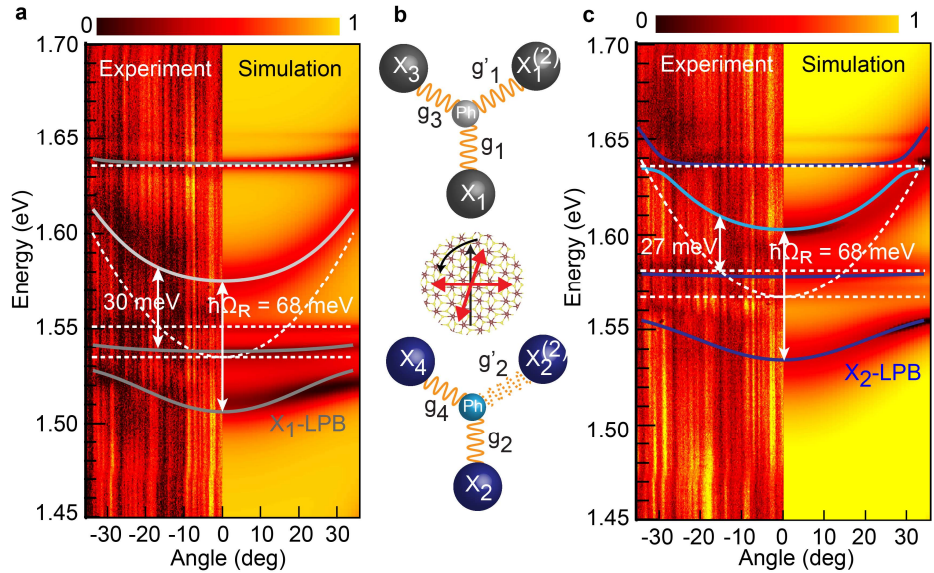

**Fig. 3. Thickness-tuned light matter coupling.** (**a**) Angle resolved reflectance (left) and transfer matrix simulation (right) for 145 nm $ReS_2$ flake on DBR, with linearly polarized incident light along $X_1$. Dispersions of the uncoupled photon mode and excitons (white dashed lines) and the four exciton-polariton modes (grey solid lines) are indicated. The uncoupled photonic mode has zero detuning with $X_1$, and the Rabi splitting between the upper and lower polariton branches is found to be 68 meV. (**b**) Spring and mass diagram representing the four-body coupled oscillator system used as a model. Incident light polarization can be used to switch between two different coupled oscillator systems. The corresponding coupling strengths ($g_1$, $g_2$, etc.) are indicated. (**c**) Angle resolved reflectance for the system with linearly polarized incident light along $X_2$. The uncoupled photonic mode has zero detuning with $X_2$ (white dashed lines). Blue solid lines indicate the polariton mode dispersions. Rabi splitting of 68 meV is observed.

In this case, there is zero detuning between the uncoupled photonic mode and $X_1$ (shown with dashed white lines). The photonic mode here couples a manifold consisting of three exciton species: $X_1$, $X_3$ and the first Rydberg excitation of $X_1$ (denoted as $X_1^{(2)}$), thus rendering this a four-body coupled oscillator system with the Hamiltonian:

$$\begin{pmatrix} E_{Ph_i}(\hbar\omega) + i\hbar\Gamma_{Ph_i} & g_i & g_j & g_i' \\ g_i & E_{Xi} + i\hbar\Gamma_{Xi} & 0 & 0 \\ g_j & 0 & E_{Xj} + i\hbar\Gamma_{Xj} & 0 \\ g_i' & 0 & 0 & E_{X_i^{(n)}} + i\hbar\Gamma_{X_i^{(n)}} \end{pmatrix}$$

Here, the subscript $i$ takes the values 1 and 2, while $j = 3$ and 4 for $X_1$ and $X_2$ polarization cases respectively. Note that $X_3$ and $X_4$ were included in this case, despite being excluded in the case of 120 nm ReS$_2$, since the photonic mode overlaps with the energies of these excitons, and the addition of ~30 layers enhances the coupling strength of $X_3$ and $X_4$ substantially according to (1) because of increased $n$. By fitting the polariton mode dispersions with the experimentally observed ones, we obtain $g_1 = 30$ meV, $g_3 = 15$ meV and $g_1' = 10$ meV. All of these interactions lie in the strong coupling regime. This is further evident from the large mode splitting of 68 meV between the lower and upper-middle polariton branches, and 30 meV between the two middle branches. Reflectance data taken at a larger angle shows Rabi splitting = 20 meV for the Rydberg UPB (see Figure S9 in the Supplemental Material).

Figure 3b shows schematically how we can also switch from one independent coupled oscillator system to another by rotating the incident linearly polarized light so that it is aligned along a different exciton dipole moment. Figure 3c shows the case when the polarized light is aligned along $X_2$. There is zero detuning between the uncoupled photon mode and $X_2$ in this case as well. We find the coupling strengths as follows: $g_2 = 30$ meV, $g_4 = 10$ meV and $g_2' = 5$ meV. All but the coupling with the Rydberg excitation satisfies the condition for strong coupling. Again, large Rabi splitting of 68 meV is achieved between the lower and upper-middle polariton branch, and 27 meV between the two middle branches.

**Revealing the bare photonic mode through temperature dependence**

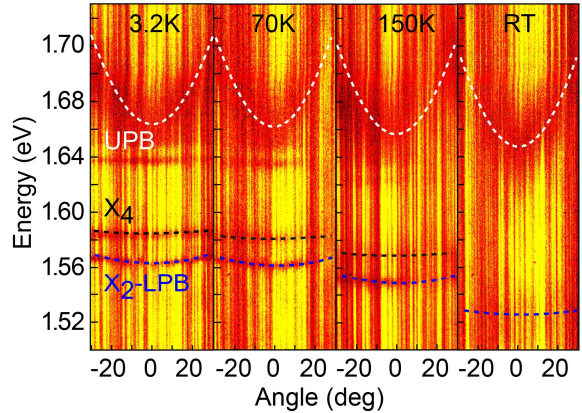

**Fig. 4. Temperature dependent detuning and coupling strength.** Angle-resolved reflectance spectra for different temperatures, with linearly polarized electric field aligned to excite only $X_2$ and $X_4$. The photon-like upper polariton branch dispersion is marked with a dashed white line. Exciton-like lower polariton branch $X_2$-LBP and $X_4$ are indicated with blue and black dashed lines respectively.

Finally, we look at temperature as an additional factor to control the degree of coupling. Figure 4 shows the results of temperature-dependent, angle-resolved reflectivity measurements on 120 nm $ReS_2$/DBR, carried out at an incident polarization angle $\theta_P$ corresponding to $X_2$ exciton. The red-shift of the exciton energy with increasing temperature in $ReS_2$ [36,49] (see Figure S10 in the Supplemental Material for fitting) can be utilized to control the extent of detuning. The photon-like UPB (marked by a dashed white line) redshifts approximately 15 meV as temperature is varied from 3.2K to RT, as a consequence of the reduced coupling strength and increased detuning. Coupling strength $g_2$ obtained from fitting falls from 32 meV to ~20 meV at 150K, while at room temperature it is negligible since the exciton population is drastically reduced. The dispersion can be treated as the bare photon mode, whose position is solely dependent on $\epsilon_b$ and the thickness of the dielectric layers. The dispersion of $X_2$-LPB decreases as well, although this effect is masked by the increasing linewidth. The absorption dip originating from the associated shoulder peak $X_4$ (marked with dashed black lines) shows negligible dispersion through all temperatures. Even though the absorption dip due to $X_4$ has been significantly enhanced in this case, it cannot be resolved beyond 150K, concurring with observations in our previous work [36]. $X_3$ (not shown) shows the same temperature-dependent behavior. Thus, temperature works as a method to control the coupling strength, the number of exciton species involved, and additionally fine-tune the polariton detuning.

## Conclusion

We have demonstrated polarization tunable strong light-matter coupling in ReS$_2$ and established it as a unique platform for creating highly anisotropic exciton-polariton manifolds. Observed tunability of the interaction as a function of incident polarization is attributed to the variation of the oscillator strength of anisotropic excitons as a function of light polarization. In addition, the dispersion of the photonic mode supported in the ReS$_2$ layer itself is polarization dependent due to optical anisotropy in the dielectric permittivity. Such systems with large oscillator strength and optical anisotropy can find applications in optoelectronics since a top mirror is not required, allowing the optically active material to be easily interfaced with electronics. The ability to switch between two independent polaritonic states via tunable light-matter coupling can see potential applications in polarization-sensitive polaritonic devices.

## Acknowledgements

S. D. acknowledges Science and Engineering Research Board (SB/S2/RJN-110/2016) Ramanujan Fellowship, SERB (CRG/2018/002845), Indian Institute of Technology Kharagpur (IIT/SRIC/ISIRD/2017-2018), and Ministry of Human Resource Development (IIT/SRIC/PHY/NTS/2017-18/75) for the funding and support for this work. A. R. C. acknowledges SERB, ECR/2017/000498 for funding for this work. D. C. acknowledges Council of Scientific and Industrial Research, JRF (09/081(1352)/2019-EMR-I) for the financial assistance. K. G. acknowledges Department of Science and Technology INSPIRE fellowship (IF180046) for the financial support.

We acknowledge D. K. Goswami and his group for the AFM measurement on the sample. We thank Joel Yuen-Zhou and S. B. N. Bhaktha for their valuable comments on this work.

Supplemental Material

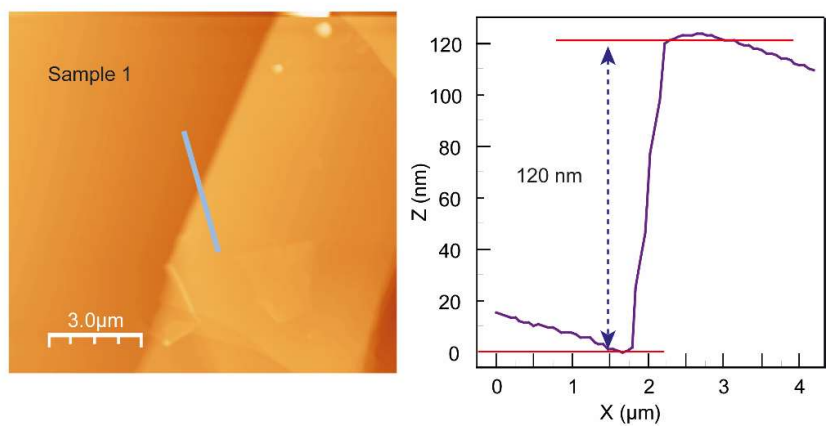

**Figure S1. Thickness measurment using AFM**. (Left) AFM image of a $ReS_2$ flake placed on DBR, (Right) Thickness along the line profile indicated on the left image with a solid line.

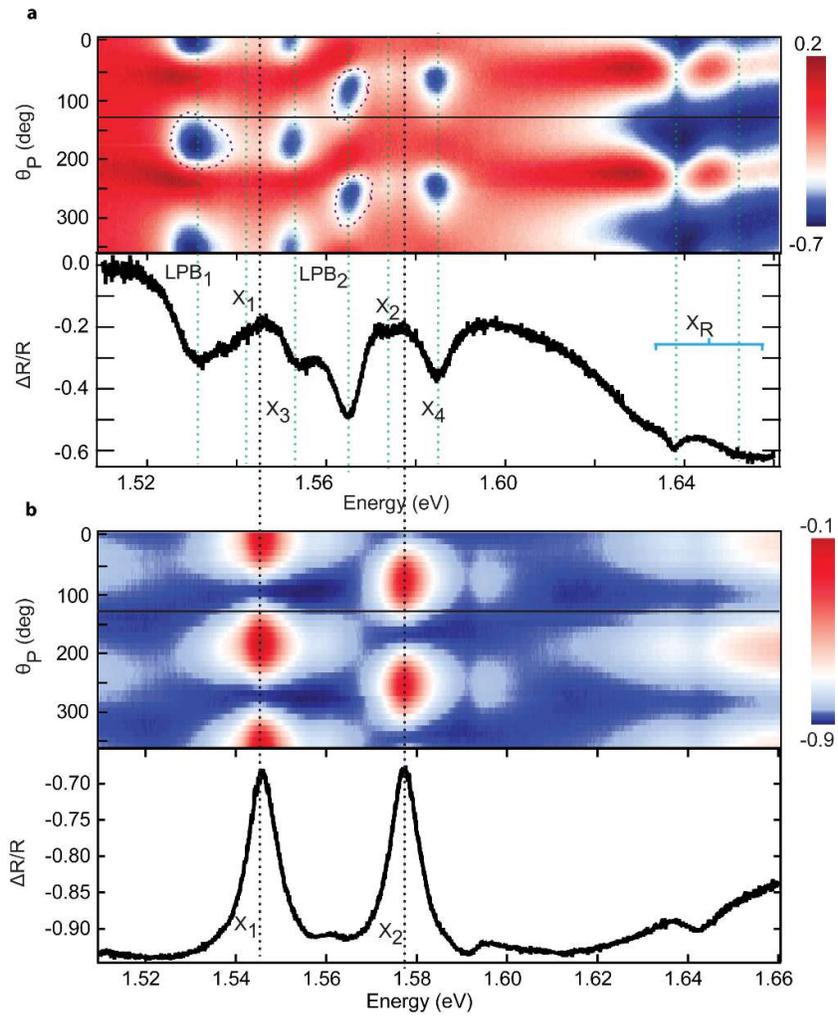

**Figure S2. Comparison between the polarization-resolved differential reflectance from ReS$_2$/DBR (top) and ReS$_2$/SiO$_2$ (bottom).** The shape of the absorption dip of excitons X$_1$ and X$_2$ in the color plot show a significant departure from ReS$_2$ on SiO$_2$/Si substrate. They appear to have distorted from an elliptical shape into a tilted-tear-drop shape, while also being red-shifted. This is taking place due to polarization-angle dependent light-matter coupling.

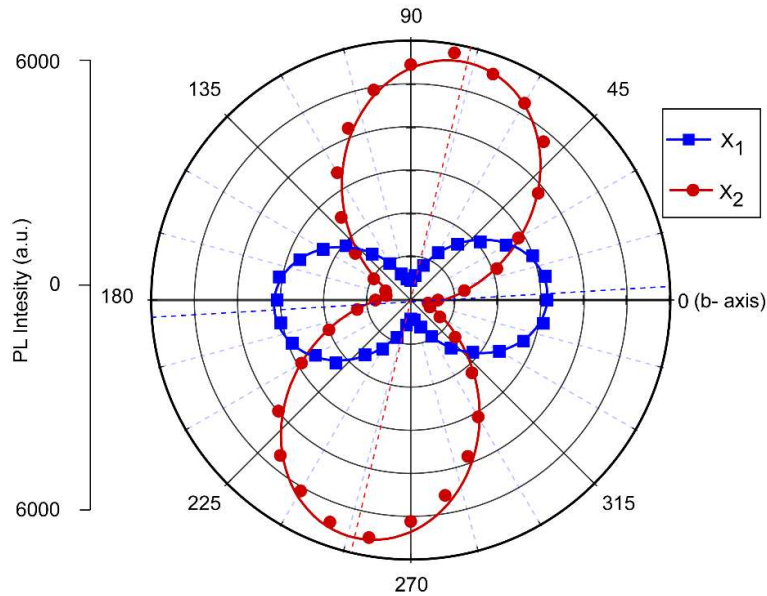

**Figure S3 | Polarization-dependent PL polar plot.** The solid coloured lines indicate the fit according to Malus' law, which determine the polarisation angle of the exciton dipole moments with respect to the b-axis.

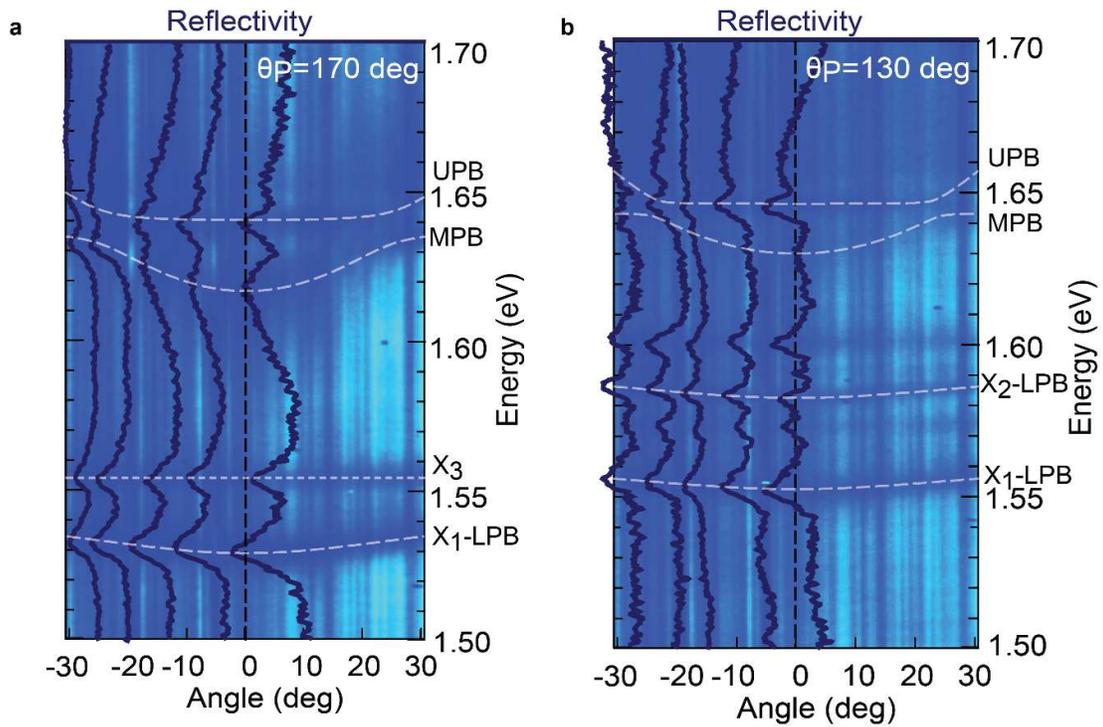

**Figure S4 | Extracting polariton dispersion using reflectivity cross-sections for 120 nm ReS₂/DBR sample.** The solid blue curves show the cross-sections of the reflectivity colour plot, for angles 0°, -10°, -20°, -25° and -30°. The fitted polariton dispersions, as obtained from the method discussed in Supporting Note 2, are shown as dashed translucent white curves.

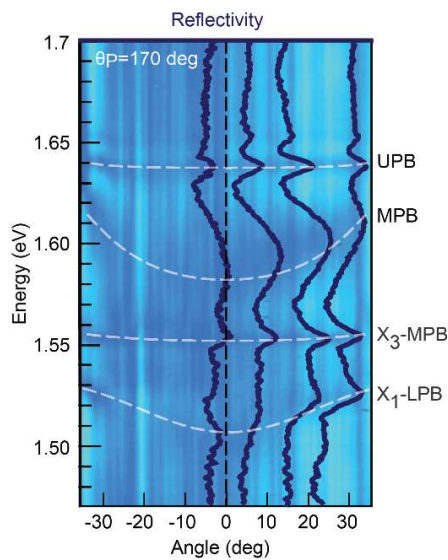

**Figure S5 | Reflectivity cross-sections for 145 nm ReS$_2$/DBR sample.** The solid blue curves show the cross-sections of the reflectivity colour plot, for angles 0°, -10°, -20°, and -30°. The fitted polariton dispersions are shown as dashed transluscent white curves.

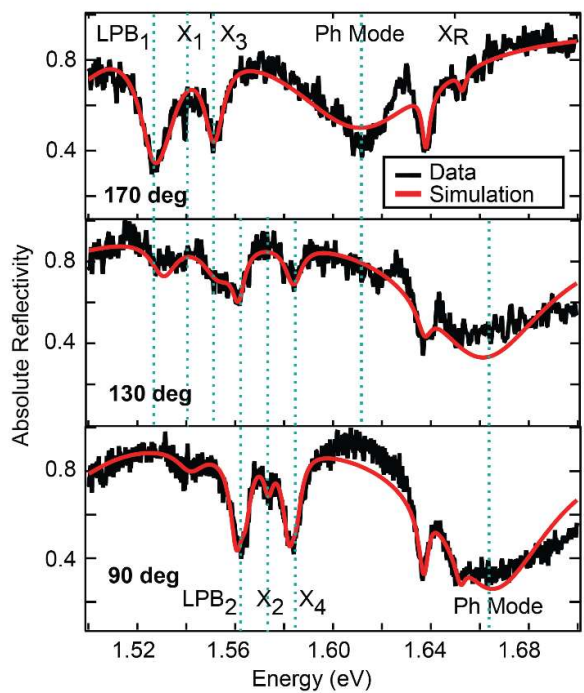

**Figure S6 | Absolute reflectivity fitting using Transfer Matrix Method.** See Supporting Note 1 for details.

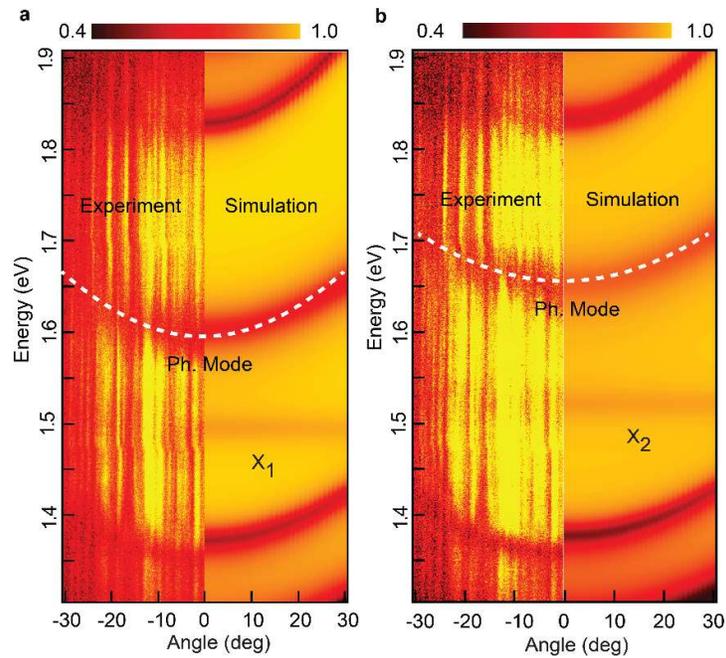

**Figure S7 | Bare photonic mode dispersions revealed at room temperature: a, b)** Angle-resolved reflectivity of the structure at room temperature for $\theta_P$ along $X_1$ and $X_2$ respectively, clearly revealing the position and linewidth of the bare photonic mode for the two different polarizations.

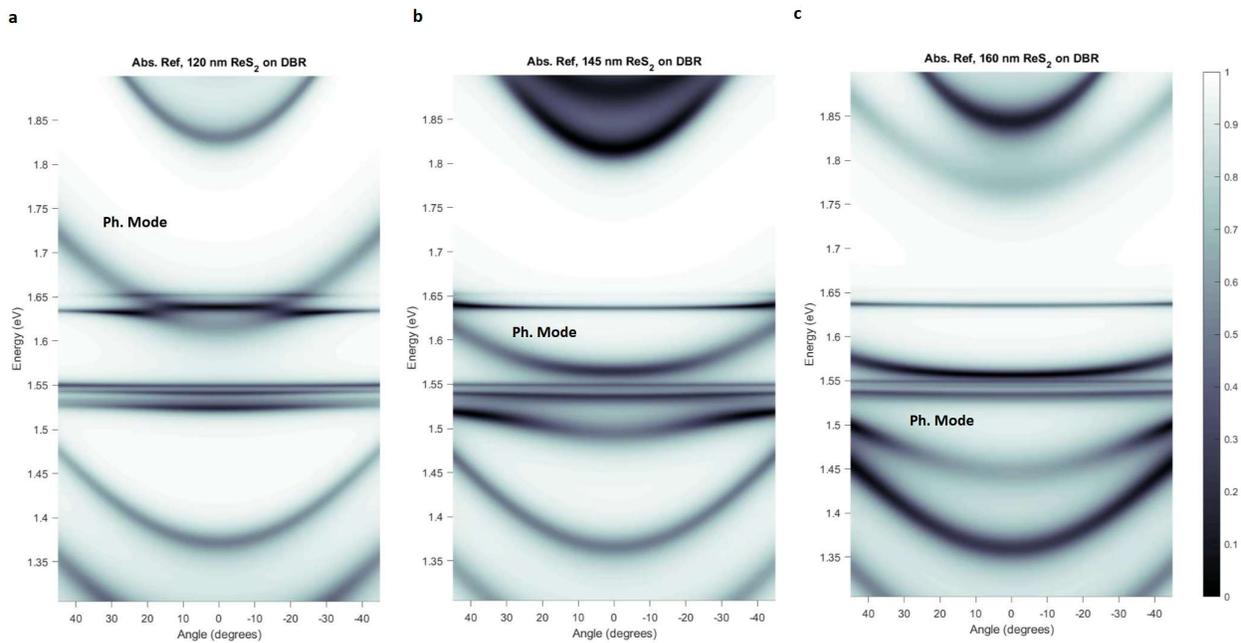

**Figure S8 | Thickness dependent detuning.** Transfer matrix simulation for reflectivity where the thickness of the top layer is increased from left to right. As the thickness is changed, the position of the photonic mode shifts. (a), (b) and (c) represent cases with positive, zero and negative detuning respectively.

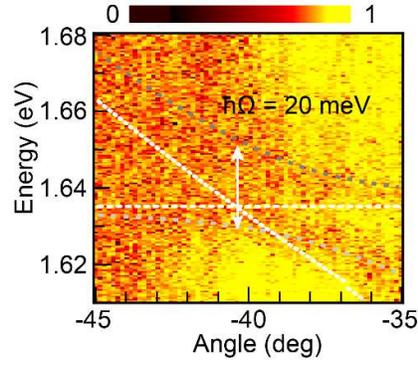

**Figure S9 | Rydberg exciton-polariton splitting in 145 nm ReS$_2$/DBR for $X_1^{(2)}$.** Angle resolved reflectance for X$_1$ polarization at a higher angle reveals VRS of 20 meV near the first Rydberg excitation. White (grey) dashed curves indicate the uncoupled (polariton) energy dispersions.

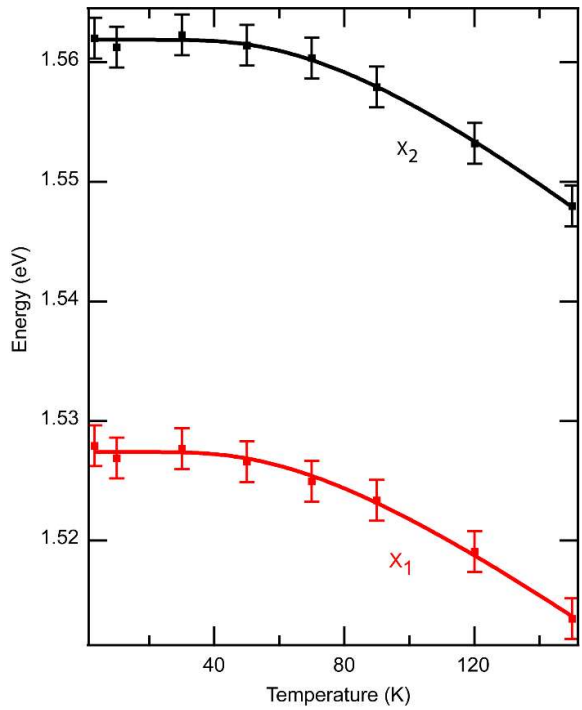

**Figure S10 | Temperature-dependent exciton peak shift.** The solid lines indicate the fit according to the O'Donnel-Chen semiconductor bandgap relation.

**Supporting Note 1: Using Transfer Matrix Method to fit reflectivity**

The absolute reflectivity of the structure at normal incidence, for three different values of $\theta_P$, is plotted in Supplementary Figure 6, along with the corresponding simulation obtained using the transfer matrix method. The dielectric function for ReS$_2$ is modelled using Lorentz oscillators to represent each exciton resonance. The top (bottom) panel shows reflectivity at $\theta_P$ for which only the excitons $X_1$ and $X_3$ ($X_2$ and $X_4$) are present. The photonic mode can be seen as a broad dip appearing in the upper half of the spectrum, overlapping and coupling with which highly enhances the exciton absorption. The photonic mode appears at different energy positions for the two values of $\theta_P$, corresponding to the two different values of the background permittivity ($\epsilon_b$) in bulk ReS$_2$. The middle panel shows the reflectivity at a particular polarization angle where six dips are visible in the exciton resonance region identified as: $X_1$-$X_4$, as well as the LPBs corresponding to exciton-polaritons formed by $X_1$ and $X_2$.

**Supporting Note 2: Extracting polariton branches from angle-resolved reflectivity**

Due to the significant noise present in the angle-resolved reflectivity data, it was not possible to extract the polariton dispersions directly. Thus, the following method was used:

i) For absolute reflectivity data corresponding to each polarization angle, 40 different cross sections were taken, each for a different emission angle ranging from -30° to 30° (see Figs. S4 and S5 for examples of such cross sections).

ii) Transfer matrix method was used as detailed in Note 1 obtain the best fit for all 40 cross sections simultaneously. The fit for one such cross section for emission angle = 0° (normal incidence) is shown in Fig S6. Thus, non-linear fitting was used to obtain the best set of parameters for a particular incident polarization by sampling a large amount of experimental data.

iii) These parameters were now used to generate the absolute reflectivity of the dielectric stack for all angles from -30° to 30°. This yields the simulated reflectivity as shown in the right side of Figs. 2(a-c) and 3(a), (c).

iv) The polariton branches are extracted from the simulated reflectivity data by tracing the minima at specified energy ranges near each branch.

**Supporting Note 3: Calculation details for coupled oscillator model**

For 120 nm $ReS_2$/DBR, at normal incidence:

| Orientation | Bare cavity position $E_{Ph_i}$ (eV) | Exciton position $E_{Xi}$ (eV) | Rydberg exciton position $X_i^{(n)}$ (eV) |
|---|---|---|---|
| **X1** | 1.600 | 1.540 | 1.637 (n=2) |
| **X2** | 1.651 | 1.573 | 1.651 (n=3) |

For 145 nm $ReS_2$/DBR, at normal incidence:

| Orientation | Bare cavity position $E_{Ph_i}$ (eV) | Exciton position $E_{Xi}$ (eV) | Additional Exciton position $E_{Xj}$ (eV) | Rydberg exciton position $X_i^{(n)}$ (eV) |
|---|---|---|---|---|
| **X1** | 1.536 | 1.536 | 1.551 | 1.638 (n=2) |
| **X2** | 1.568 | 1.568 | 1.571 | 1.635 (n=2) |